\documentclass[preprint]{emulateapj}
\usepackage{amsmath,amssymb,graphicx,longtable,booktabs}

\newcommand{\magarcsec}{mag arcsec$^{-2}$}	
\newcommand{\king}{KINGPHOT}
\newcommand{\ug}{$u^\ast-g^\prime$}
\newcommand{\gi}{$g^\prime-i^\prime$}
\newcommand{\gz}{$g^\prime-z^\prime$}
\newcommand{\iz}{$i^\prime-z^\prime$}
\newcommand{\objname}{M59-UCD3}
\newcommand{\sersic}{Sersi\'c}
\newcommand{\kms}{${\rm km\ s^{-1}}$}

\defcitealias{Strader13}{S13}
\defcitealias{Chilingarian08}{CM08}
\defcitealias{Norris14}{N14}

\shorttitle{The most massive UCD in Virgo}
\shortauthors{Liu, Peng et al.}

\begin{document}

\title{The Most Massive Ultra-Compact Dwarf Galaxy in the Virgo Cluster}

\author{Chengze Liu,\altaffilmark{1,2}}
\author{Eric W. Peng,\altaffilmark{3,4,5}}
\author{Elisa Toloba\altaffilmark{6,7}}
\author{J. Christopher Mihos\altaffilmark{8}}
\author{Laura Ferrarese\altaffilmark{9}}
\author{Karla Alamo-Mart\'{i}nez\altaffilmark{10,11,12,13}}
\author{Hong-Xin Zhang\altaffilmark{14,10,12,13}}
\author{Patrick C\^ot\'e\altaffilmark{9}}
\author{Jean-Charles Cuillandre,\altaffilmark{15}}
\author{Emily C.\ Cunningham\altaffilmark{6}}
\author{Puragra Guhathakurta\altaffilmark{6}}
\author{Stephen Gwyn,\altaffilmark{9}}
\author{Gregory Herczeg\altaffilmark{5}}
\author{Sungsoon Lim,\altaffilmark{4,5}}
\author{Thomas H.\ Puzia\altaffilmark{10}}
\author{Joel Roediger\altaffilmark{9}}
\author{Rub\'en S\'anchez-Janssen\altaffilmark{9}}
\author{Jun Yin\altaffilmark{16}}

\altaffiltext{1}{Center for Astronomy and Astrophysics, Department of Physics and Astronomy, Shanghai Jiao Tong University, Shanghai 200240, China}
\altaffiltext{2}{Shanghai Key Lab for Particle Physics and Cosmology, Shanghai Jiao Tong University, Shanghai 200240, China}
\altaffiltext{3}{Corresponding author: peng@pku.edu.cn}
\altaffiltext{4}{Department of Astronomy, Peking University, Beijing 100871, China}
\altaffiltext{5}{Kavli Institute for Astronomy and Astrophysics, Peking University, Beijing 100871, China}
\altaffiltext{6}{UCO/Lick Observatory, University of California, Santa Cruz, 1156 High Street, Santa Cruz, CA 95064, USA}
\altaffiltext{7}{Texas Tech University, Physics Department, Box 41051, Lubbock, TX 79409-1051, USA}
\altaffiltext{8}{Department of Astronomy, Case Western Reserve University,10900 Euclid Ave, Cleveland, OH 44106, USA}
\altaffiltext{9}{Herzberg Institute of Astrophysics, National Research Council of Canada, Victoria, BC V9E 2E7, Canada}
\altaffiltext{10}{Departamento de Astronom\'ia y Astrof\'isica, Pontificia Universidad Cat\'olica de Chile, 7820436 Macul, Santiago, Chile}
\altaffiltext{11}{FONDECYT Postdoctoral Fellow}
\altaffiltext{12}{CAS-CONICYT Fellow}
\altaffiltext{13}{Chinese Academy of Sciences South America Center for Astronomy, Camino EI Observatorio \#1515, Las Condes, Santiago, Chile}
\altaffiltext{14}{National Astronomical Observatories, Chinese Academy of Sciences, Beijing 100012, China}
\altaffiltext{15}{CEA/IRFU/SAp, Laboratoire AIM Paris-Saclay, CNRS/INSU, Universit\'e Paris Diderot, Observatoire de Paris, PSL Research University, F-91191 Gif-sur-Yvette Cedex, France}
\altaffiltext{16}{Key Laboratory for Research in Galaxies and Cosmology, Shanghai Astronomical Observatory, Chinese Academy of Sciences, 80 Nandan Road Shanghai, 200030 China}

\begin{abstract}
We report on the properties of the most massive ultra-compact dwarf galaxy (UCD) in the nearby Virgo Cluster of galaxies using imaging from the Next Generation Virgo Cluster Survey (NGVS) and spectroscopy from Keck/DEIMOS. This object (M59-UCD3) appears to be associated with the massive Virgo galaxy M59 (NGC 4621), has an integrated velocity dispersion of 78~\kms, a dynamical mass of $3.7\times10^8 M_\odot$, and an effective radius ($R_e$) of 25~pc. With an effective surface mass density of $9.4\times10^{10} M_\odot\ {\rm kpc}^{-2}$, it is the densest galaxy in the local Universe discovered to date, surpassing the density of the luminous Virgo UCD, M60-UCD1. \objname\ has a total luminosity of $M_{g'}=-14.2$~mag, and a spectral energy distribution consistent with an old (14~Gyr) stellar population with [Fe/H]$=0.0$ and $\rm{[\alpha/Fe]}=+0.2$. We also examine deep imaging around M59 and find a broad low surface brightness stream pointing towards \objname, which may represent a tidal remnant of the UCD progenitor. This UCD, along with similar objects like M60-UCD1 and M59cO, likely represents an extreme population of tidally stripped galaxies more akin to larger and more massive compact early-type galaxies than to nuclear star clusters in present-day dwarf galaxies.

\end{abstract}

\keywords{galaxies: clusters: individual (Virgo) --- galaxies: dwarf --- galaxies: evolution --- galaxies: individual (M59) --- galaxies: nuclei --- galaxies: star clusters: general}


\section{Introduction}
\label{sec:intro}
\setcounter{footnote}{0}

The classic separation between galaxies and star clusters based on properties such as size and luminosity has become increasingly blurred since the discovery of a family of compact stellar systems \citep{Hilker99,Drinkwater00} that are intermediate between compact elliptical galaxies (e.g., M32) and the most massive globular star clusters. These ``ultra-compact dwarfs'' (UCDs), or ``dwarf-globular transition objects'' (DGTOs, \citealt{Hasegan05}), have sizes $10<r_h<100$~pc and luminosities of $M_V<-9$~mag, and are of indeterminate origin. Their properties are similar to both the nuclei of low-mass galaxies \citep{Georgiev14}, and to massive globular clusters (GCs), suggesting possible connections to both of those populations \citep[see, e.g.,][for a review of current ideas]{Brodie11,Norris14,Zhang15}.
UCDs have so far been found mostly in dense environments, at the centers of galaxy clusters, or near massive galaxies, implying that environmental effects, such as tidal stripping, may play an important role in their formation \citep[e.g.,][]{Bekki03,Pfeffer13}. It is also possible that UCDs formed as compact galaxies at early times when the Universe was much denser \citep{Milosavljevic14}, or are simply the massive extension of the GC population \citep{Mieske02, Mieske12}, although UCD kinematics show that the latter is probably not the case in M87 \citep{Zhang15}.

In recent years, two very massive UCDs have been discovered in the nearby Virgo Cluster. These UCDs, M60-UCD1 \citep[][hereafter S13]{Strader13} and M59cO \citep[][hereafter CM08]{Chilingarian08} are among the densest stellar systems in the local Universe, comparable to the densest stellar nuclei of local galaxies. M60-UCD1 had, up until now, the highest density of any known galaxy.  
The recent discovery of a massive black hole in M60-UCD1 \citep{Seth14} raises the possibility that UCDs could be a common host of galactic nuclear black holes (\citealt{Mieske13}; although not necessarily ubiquitous, {\it cf.} \citealt{Frank11}). The black hole in M60-UCD1 has a high mass fraction ($\sim0.15$), two orders of magnitude larger than typical values \citep{volonteri10}, which suggests a tidal stripping origin. 

In this paper, we report on the properties of \objname, a remarkable UCD associated with the Virgo Cluster early-type galaxy M59 (NGC~4621/VCC~1903). While this paper was in the late stages of preparation, \citet{Sandoval15} reported a concurrent discovery of \objname\ using imaging from the SDSS and Subaru telescopes. Our paper uses independent imaging, and is based on follow-up spectroscopy obtained in 2014 and 2015.

\section{Observations}
\label{sec:obs}

\subsection{Imaging}
\label{sec:imaging}

This study is part of a systematic search for UCDs across the entire Virgo Cluster using the homogeneous imaging of the Next Generation Virgo Cluster Survey (NGVS; \citealt{Ferrarese12}).  
The NGVS imaged the Virgo cluster out to the virial radii of its
two main subclusters, for a total sky coverage of 104 deg$^2$,
in four optical bands ($u^*g'i'z'$). The NGVS produced two sets of stacked images for each $1^\circ\times1^\circ$ field: a ``long exposure'' stack consisting of images with exposure times from 2055 to 6400 seconds (depending on the band), and a ``short exposure'' stack, consisting of images with exposure times from 40 to 250 seconds. The long exposure images saturate at $g'\sim18.5$~mag for point sources, so we use the short exposure images to study compact galaxies and galaxy nuclei. The methods used to generate catalogs and aperture photometry are similar to those described in \citet{Durrell14}.

We selected UCD candidates based on their colors in the $($\ug$)-($\gz$)$ diagram, then fit them with \citet{King66} models convolved with the point spread function (PSF) using the \king\ software package \citep{Jordan05}. The King model fits allow us to determine which objects are extended in appearance.

\begin{figure}
\epsscale{1.15}
\plotone{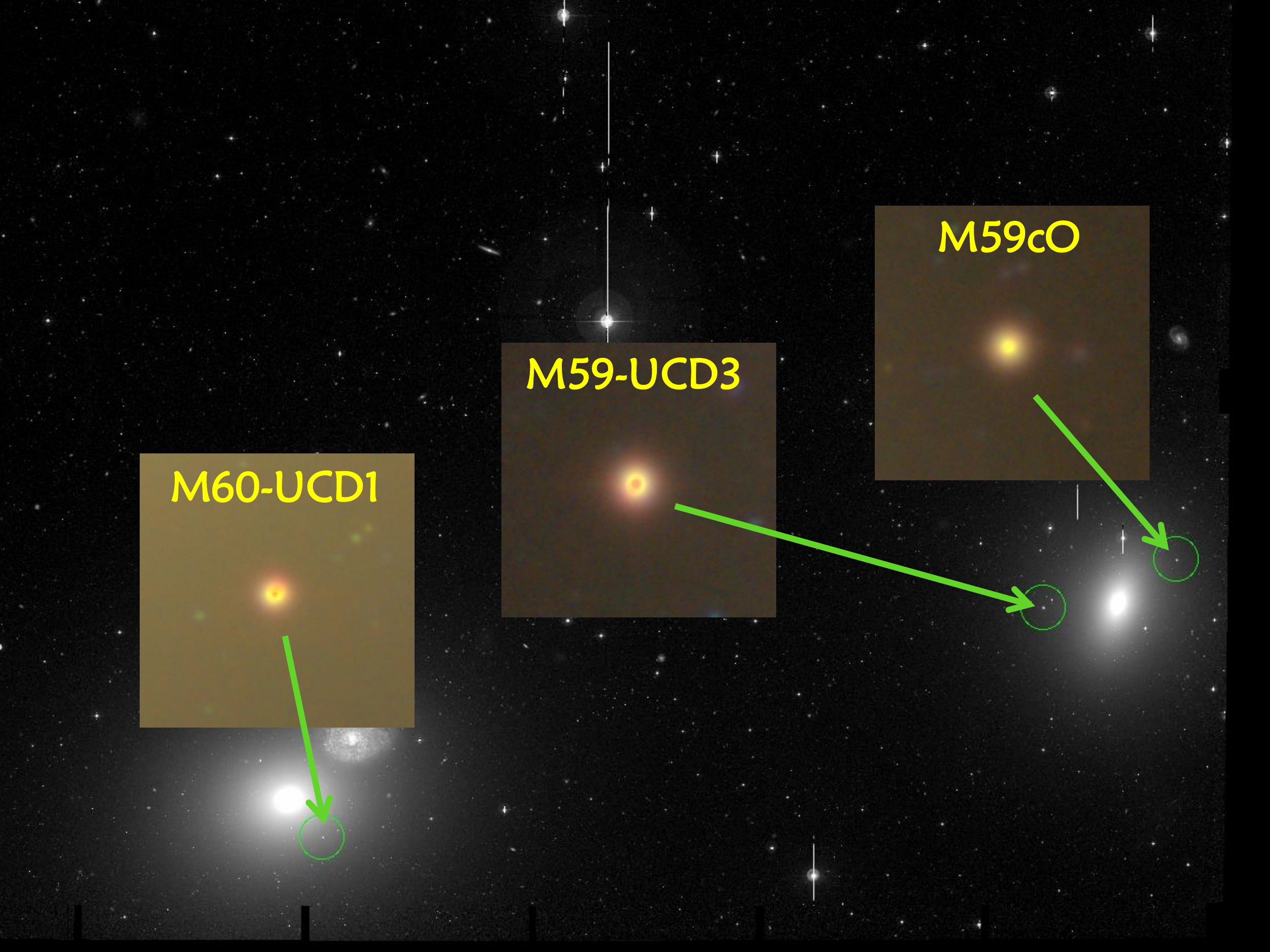}
\caption{NGVS images of the three brightest UCDs in the Virgo cluster. From left to right: M60-UCD1 \citepalias{Strader13}, \objname\ (this work) and M59cO \citepalias{Chilingarian08}.}
\label{fig:bright_ucds}
\end{figure}

We identified NGVSJ124211.05+113841.24 (\objname) as a bright, extended object $130\arcsec$ to the east of the massive early-type galaxy M59 (9.4~kpc at the distance of M59). There is no HST imaging of \objname,
so our photometric analysis is based solely on the NGVS images, which had image quality of ${\rm FWHM(}g'{\rm)}=0\farcs58$ and ${\rm FWHM(}i'{\rm )}=0\farcs46$ ($R_e$ of 12.0 and 9.5~pc, respectively, at M59).
Figure~\ref{fig:bright_ucds} shows \objname\ relative to the massive early-type galaxies, M60 (left) and M59 (right), and the other bright UCDs, M60-UCD1 \citepalias{Strader13} and M59cO \citepalias{Chilingarian08}.

\subsection{Spectroscopy}
\label{sec:spec}

The radial velocity of M59-UCD3 was measured as 440~\kms\ in a 320 second exposure obtained on 29 November 2014 with the Supernovae Integral Field Spectrograph (SNIFS) on the UH 2.2m telescope \citep{Lantz04}. This confirmed that \objname\ is both in the Virgo Cluster, and likely to be associated with M59 ($V_{hel}=467\pm5$~\kms, \citealt{Cappellari11}). The distance to M59 is $14.9\pm0.5$~Mpc \citep{Mei07, Blakeslee09}, and for the rest of the paper, we assume that \objname\ is at the same distance. 

To confirm the SNIFS radial velocity and obtain a measure of the internal velocity dispersion of \objname, on 14 April 2015, we obtained a 300 second exposure with the Keck/DEIMOS spectrograph \citep{Faber03} using the 600-line grating and the $0\farcs7$-wide long slit, giving a spectral range of $\lambda\lambda$4800--9500${\rm \AA}$, a spectral resolution of $2.8{\rm \AA}$ (corresponding to $\sigma=50$~\kms), and peak signal-to-noise of 107 ${\rm \AA}^{-1}$. These data were reduced using the \citet{SimonGeha07} modification of the DEIMOS {\tt spec2d} pipeline \citep{Cooper12,Newman13}, 
and the spectrum was optimally extracted by fitting a Gaussian function (${\rm FWHM}=10$~pixels or $1\farcs85$) to the spatial intensity profile.

\section{Results}
\label{sec:results}

\subsection{Photometry and Structural Parameters}
\label{sec:results_im}

\begin{deluxetable*}{lccccccccc}
\tablewidth{0pt}
\tablecaption{Photometric properties of Luminous Virgo Cluster UCDs\label{tab:phot}}
\tablehead{
\colhead{Name} &
\colhead{RA(J2000)} &
\colhead{Dec(J2000)} &
\colhead{$g^\prime$} &
\colhead{\ug} &
\colhead{\gi} &
\colhead{\iz} &
\colhead{$R_{e,NGVS}$} &
\colhead{$R_{e,HST}$} &
\colhead{$\langle\mu_{g'}\rangle_e$} \\
\colhead{} &
\colhead{} &
\colhead{} &
\colhead{mag} &
\colhead{mag} &
\colhead{mag} &
\colhead{mag} &
\colhead{pc} &
\colhead{pc} &
\colhead{\magarcsec} \\
\colhead{(1)} &
\colhead{(2)} &
\colhead{(3)} &
\colhead{(4)} &
\colhead{(5)} &
\colhead{(6)} &
\colhead{(7)} &
\colhead{(8)} &
\colhead{(9)} &
\colhead{(10)}
 \\
}

\startdata

\objname$^\ast$ & 190.5460471 & 11.6447937 & 16.74 & 1.63 & 1.10 & 0.28 & $25\pm2$& \nodata & 16.42 \\
M60-UCD1 & 190.8998699 & 11.5346389 & 17.49 & 1.77 & 1.13 & 0.27 & $36\pm2$ & 27 & 17.72 \\
M59cO    & 190.4805689 & 11.6677212 & 17.93 & 1.70 & 1.11 & 0.27 & $37\pm2$ & 35 & 18.50 \\

\enddata

\tablenotetext{4}{Total $g^\prime$ magnitude from a single-\sersic\ fit to \objname\ and M59cO, and a double-\sersic\ fit to M60-UCD1.}
\tablenotetext{5--7}{Colors measured within a $3\arcsec$ diameter aperture.}
\tablenotetext{8}{Effective radius and uncertainty measured from profile fits in the NGVS image, in parsecs, assuming distances of $14.9\pm0.5$~Mpc for M59 UCDs and $16.5\pm0.6$~Mpc for M60 \citep{Blakeslee09}. Size uncertainties include the fitting uncertainty and the distance uncertainty added in quadrature, but not systematic errors (PSF, galaxy subtraction) that could be at the level of 20\%. For M60-UCD1, $R_e$ contains half the total light from a double-\sersic\ fit.}
\tablenotetext{9}{Effective radius measured using HST imaging, from \citet{Norris14}} 
\tablenotetext{10}{Mean $g'$-band surface brightness within $R_e$, assuming the effective radius in Column (8).}
\tablenotetext{$\ast$}{\citet{Sandoval15} reported $g_{SDSS}=16.81$~mag, $(g-i)_{SDSS}=1.20$~mag and $R_{e,Subaru}=20\pm4$~pc, all of which are consistent with our measurements.
}
\end{deluxetable*}

To provide an internally consistent comparison of the photometric properties of \objname\ with M60-UCD1 and M59cO, we used NGVS data for all three objects in the same way (Table~\ref{tab:phot}). In the $g'$-band, we find that \objname\ is 0.75~mag brighter than M60-UCD1, and 1.19~mag brighter than M59cO. \objname\ has the highest effective surface brightness of the three UCDs, with $\langle\mu_{g'}\rangle_e=16.42$~\magarcsec. After correcting for foreground extinction \citep{Schlafly11}, the total luminosity of \objname\ is $M_{g'}=-14.2$~mag, or $L_{g'}=5.3\times10^7 L_\odot$.

\begin{figure}
\epsscale{1.15}
\plotone{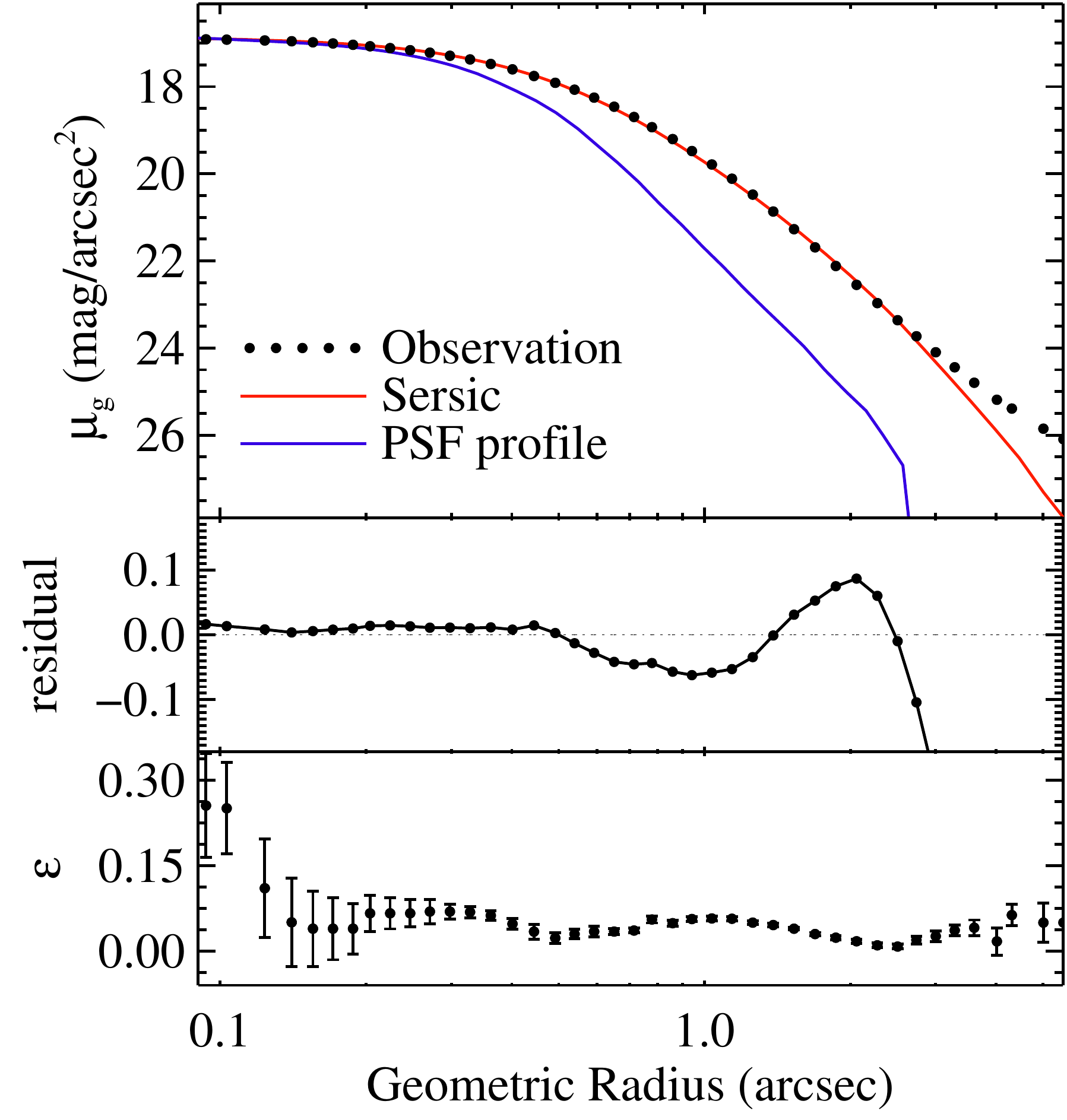}
\caption{{\it Top:} The $g'$-band surface brightness profile (dots) with the PSF (blue) and best-fit PSF-convolved \sersic\ (red) profiles overplotted. {\it Middle:} Residuals from the \sersic\ fit. {\it Bottom:} The best fit ellipticity profile. All are plotted as a function of the geometric radius $a(1-\epsilon)^{1/2}$, in which $a$ is the semi-major axis.}
\label{fig:profile}
\end{figure}

We fit the $g'$-band surface brightness profile of \objname\ with a PSF-convolved \sersic\ profile (Figure~\ref{fig:profile}). Beyond $R\approx 1\farcs2$, the data show a slight excess above a \sersic\ model. 
The uncertainty in subtracting the light from M59 makes it difficult to ascertain whether \objname\ is better fit by a double-\sersic, so we only report results for a single-\sersic\ fit. Unlike for M60-UCD1, \objname's ellipticity does not vary much with radius. The object is fairly round, except at the very center where systematic PSF uncertainties dominate. \objname\ is best fit by a \sersic\ profile with effective radius, $R_e=0\farcs345\pm0\farcs03$ ($25\pm2$~pc), \sersic\ index, $n=2.5$, and surface brightness at $R_e$, $\mu_e(g')=17.57$~\magarcsec.

Table~\ref{tab:phot} includes $R_e$ from HST imaging for M60-UCD1 and M59cO from \citet[][hereafter N14]{Norris14}. The size we measure for M60-UCD1 ($36\pm2$~pc) is larger than that measured in HST/ACS images (27~pc), although our measurement for M59cO is consistent with \citetalias{Norris14}. Our $R_e$ estimates may be biased toward larger sizes, given the larger PSF in the NGVS data \citep[see, e.g.,][]{Puzia14}, or $R_e$ may be affected by the details of subtracting light from M60. This is why we perform a relative comparison of all three UCDs using the same images and same fitting software, as relative measurements are more reliable. Even if we adopt the smaller size value for M60-UCD1, it would still have a lower effective surface brightness than \objname.

\subsection{Velocity and Internal Kinematics}
\label{sec:results_spec}

Using our Keck/DEIMOS spectrum, we measured the line-of-sight velocity and velocity dispersion with the penalized pixel fitting (pPXF) software of \citet{Cappellari04}, which fits a broadened combination of stellar templates with different weights. We used 31 stellar templates with spectral types from B to M, observed previously with DEIMOS using the same grating. Because sources can be mis-centered on the slit, we applied a velocity correction using the atmospheric A and B absorption bands, whose centers depend on how the source illuminates the slit. The heliocentric radial velocity of \objname\ is $V_r=447\pm3$~\kms, and the integrated velocity dispersion is $\sigma=77.8\pm1.6$~\kms. As a consistency check, we also measured the velocity dispersion of M59cO, which we observed simultaneously with \objname\ through the same slit. Our measurement of $\sigma=28.1\pm4.2$~\kms\ is consistent with that of \citetalias{Norris14}, who reported a value of $\sigma=29.0\pm2.5$~\kms.

We estimate the mass of \objname\ using a spherical isotropic Jeans equation analysis. We deproject the best-fit \sersic\ profile, assume mass follows light, derive the projected dispersion profile, convolving both the model dispersion and luminosity profiles with a Moffat PSF (FWHM$=1\farcs2$). We then apply a $0\farcs7$-wide slit to our model to simulate the observed integrated velocity dispersion. Matching our observed dispersion gives a total dynamical mass of $(3.7\pm0.4)\times10^8 M_\sun$, with a mass-to-light ratio of $M_{dyn}/L_{g'}=7.1\pm0.7$, or $M_{dyn}/L_V=4.9\pm0.5$ assuming $g'-V=0.40$~mag \citep[using the transformation in][]{Peng06b}.
Our model gives an average velocity dispersion within $R_e$ of $\sigma_e=94.1\pm1.9$~\kms.

\subsection{Stellar Populations}
\label{sec:result_ssp}

\begin{figure*}
\epsscale{0.9}
\plotone{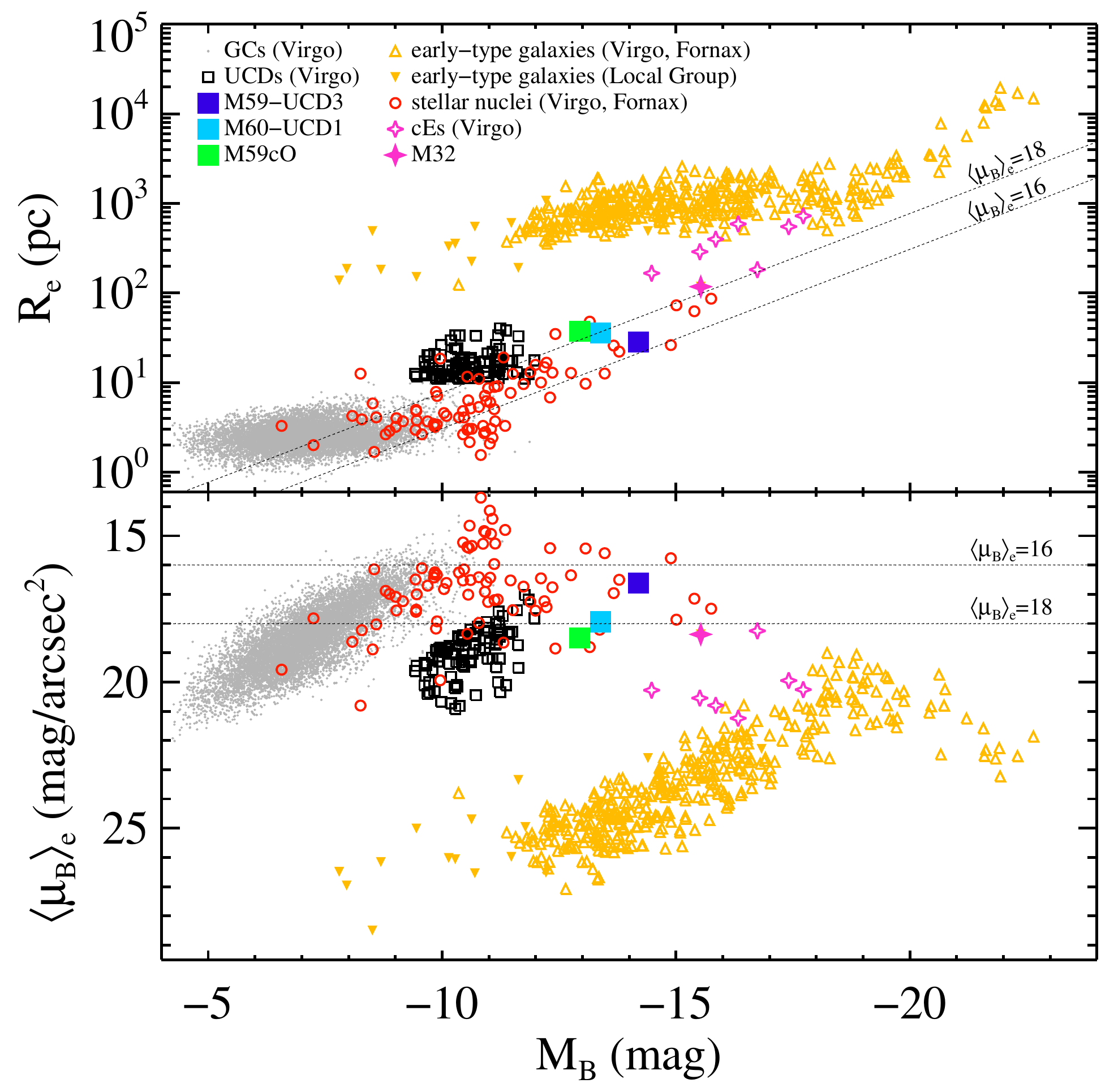}
\caption{The relationship between size ($R_e$, {\it top}), effective surface brightness ($\langle\mu_B\rangle_e$, {\it bottom}) and luminosity ($M_B$) for dynamically hot stellar systems, mostly in the Virgo or Fornax Clusters. Dotted lines denote constant effective surface brightness $\langle\mu_B\rangle_e=16,~18$~\magarcsec. We plot cluster early-type galaxies (ETGs) and Local Group ETGs from the compilation described in \citet{Ferrarese12}, compact ellipticals \citep{Guerou15}, M32 \citep{Cappellari06}, galactic stellar nuclei \citep{Cote06,Turner12}, M87 UCDs \citep [defined to have $R_e>10$~pc,][]{Zhang15}, GCs \citep{Jordan09}, and the three Virgo UCDs discussed in this paper.}
\label{fig:rh_mag}
\end{figure*}

We compared the absorption line indices of \objname\ with single stellar population (SSP) models to derive age, metallicity, [$\alpha/$Fe], and a stellar $M/L$. We used a \citet{Kroupa01} initial mass function (IMF) and the LIS-5\AA\ system \citep{Vazdekis10} where we put the models and the target spectrum at a constant resolution of 5\AA. By comparing the H$\beta$ and ${\rm [MgFe]^\prime}$ indices with the SSP models of \citet{Vazdekis15} we obtain age and metallicity estimates of 14~Gyr and [Fe/H]$=-0.01\pm0.21$. The age is fixed to the oldest model available because H$\beta$ is outside the model grid. This makes M59-UCD3 comparable in age and metallicity to the oldest and most metal-rich dE nuclei, and more metal-rich than typical dEs \citep[e.g.][]{Michielsen08,Toloba14}. We compare the Mgb and $\langle{\rm Fe}\rangle$ indices to 14~Gyr SSP models to obtain [$\alpha/$Fe]$=+0.21\pm0.07$.
For these SSP parameter values, the models provide a stellar $M_\star/L_V=3.4\pm0.9$ for a Kroupa IMF or $5.7\pm1.5$ for a Salpeter IMF. 
\objname\ has a dynamical $M_{dyn}/L$ between the $M_\star/L$ values derived from SSP models using Kroupa and Salpeter IMFs, and could be consistent with either, within the uncertainties.

\subsection{Comparison with \citet{Sandoval15}}

\citet{Sandoval15} reported the discovery of \objname\ using SDSS and Subaru imaging, and spectroscopy from SOAR/Goodman. They reported $g,r,i$ photometry, a size of $R_e=20\pm4$~pc, a radial velocity $V_r=373\pm18$~\kms, and SSP age and abundances.  The photometry between our two studies for common filters ($g$ and $i$) is consistent, as are the sizes, and the derived values of [Fe/H] and [$\alpha/$Fe]. Our age estimate (14~Gyr) is larger than theirs ($8.6\pm2.2$~Gyr), but the model grids at old ages are  closely spaced and this discrepancy is not alarming, although merits further investigation. Our radial velocity is 74~\kms\ higher, a significant difference that is difficult to explain given the high $S/N$ of both of our observations. \citet{Sandoval15} do not report a velocity dispersion.

\section{Discussion}
\label{sec:discussion}
\subsection{M59-UCD3, an Extreme Spheroidal System}

\objname\ is the most extreme object of its type discovered to date, although it is basically similar to M60-UCD1. Figure~\ref{fig:rh_mag} shows the location of dynamically hot stellar systems, mostly in the Virgo Cluster, in the parameter space of size ($R_e$), effective $B$-band surface brightness ($\langle\mu_B\rangle_e$), and $B$-band luminosity ($M_B$). The three luminous UCDs in M59 and M60 are more luminous than other UCDs in Virgo, but they are significantly more compact than objects categorized as ``compact ellipticals'' (cEs), which typically have $R_e\gtrsim100$~pc. Lines of constant $\langle\mu_B\rangle_e$ show that \objname\ is one of the densest galaxies known. These luminous UCDs appear to bridge the parameter space between massive star clusters and compact galaxies, further blurring the boundary between the two classes. The only objects in Virgo that approach or exceed the effective surface density of \objname\ are nuclear star clusters (which may have brighter $B$-band fluxes due to rejuvenated stellar populations) and some of the most massive GCs. The total effective surface mass density ($\Sigma_{eff}$, mass density within $1R_e$) is $9.4\times10^{10} M_\odot\ {\rm kpc}^{-2}$, which is roughly the maximum $\Sigma_{eff}$ observed anywhere in the Universe. As discussed in \citet{Hopkins10}, finding these extremely dense systems can illuminate the physical feedback processes that limit the density of stellar systems. Finding more UCDs in other environments \citepalias[e.g.,][]{Norris14} will help disentangle the relationships between these families of objects.

\subsection{Tidal Debris}

\begin{figure}
\epsscale{1.0}
\plotone{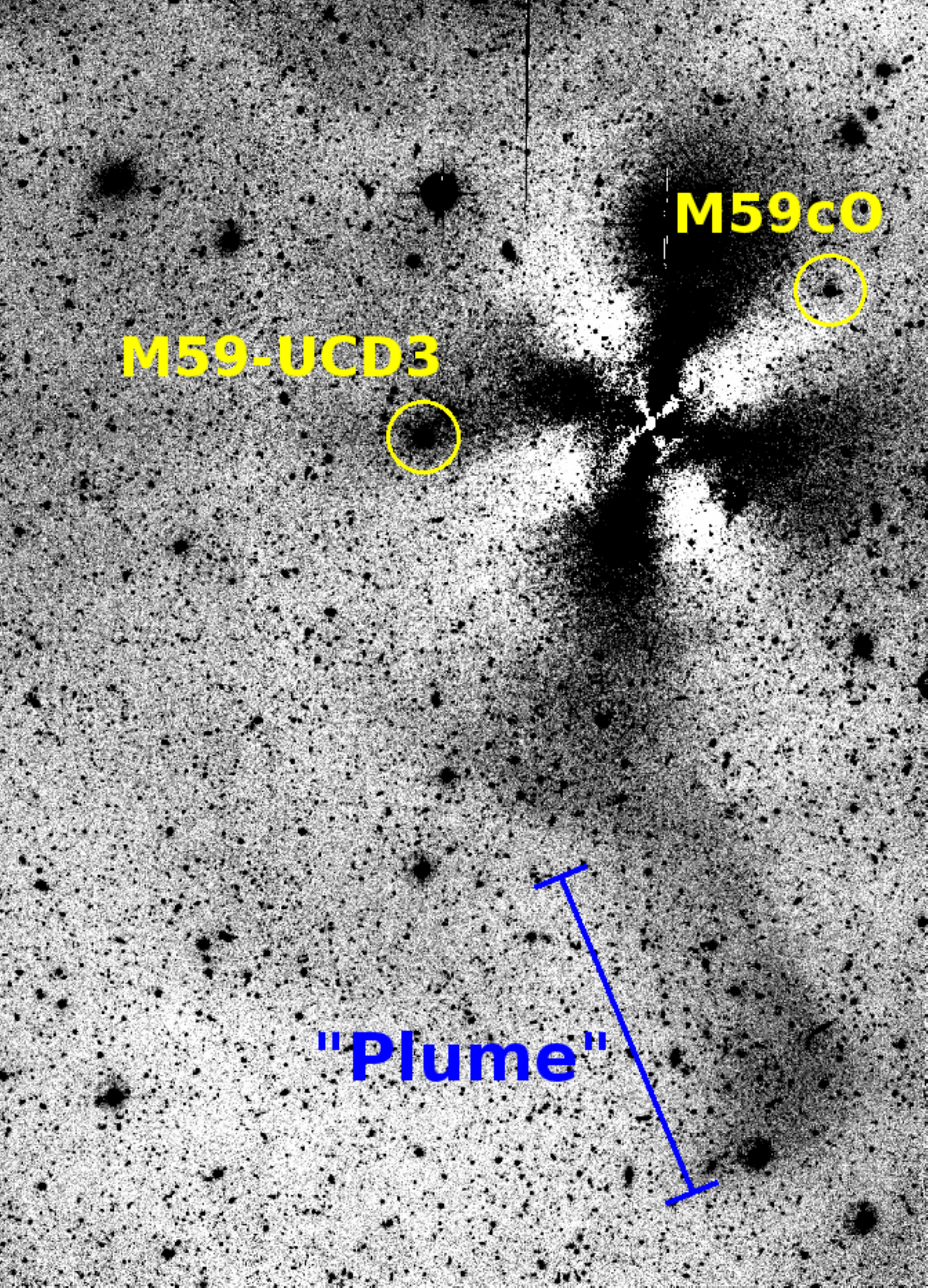}
\caption{The M59 region with an ellipse model of M59 subtracted, with north up, east to the left. The image was binned $3\times3$ pixels ($0\farcs56\times0\farcs56$) to enhance low surface brightness features. The "X" pattern is the residual of the model subtraction. \objname\ is circled to the left (east) of M59, and M59cO is to the upper right of M59. Below M59 (south) is a broad linear plume that we have identified as a potential stellar stream in M59's halo. The length of the blue line is $200\arcsec$ (14.4~kpc). The stream points roughly in the direction of \objname, and could be related to \objname's parent body.}
\label{fig:stream}
\end{figure}

If UCDs are the compact remnants of relatively recent tidal stripping events, we may find stripped stellar material in their vicinities. There has been little evidence of tidal debris linked to UCDs (with a few exceptions, {\it cf.}\ \citealt{Foster14,Mihos15}), but simulations show that the window of time during which tidally stripped starlight would be visible with current facilities is relatively short \citep{Rudick09}. Given the high mass and metallicity of \objname, however, any stripped progenitor could be very massive, leaving a significant amount of debris. The average present-day stellar mass fraction of nuclear star clusters in early-type galaxies is of order $4\times10^{-3}$ \citep[e.g.,][]{Turner12}. If this was true for a system that once hosted \objname, then the progenitor galaxy would have had a stellar mass of $\sim10^{10} M_\odot$, or about 15\% the mass of M59. This logic, however, is uncertain in a few ways: 1) The $1\sigma$ scatter in nuclear mass fraction is a factor $\sim10$, 2) This fraction is largely defined by lower mass nuclei, and 3) Present-day UCDs may not be equivalent to today's nuclear star clusters, but may instead combine a nuclear component with a stripped bulge-like component. 

We searched for fine structure and tidal streams around M59 that could be associated with either \objname, or M59cO. We used the long exposure NGVS $g'$-band image reduced with the Elixir-LSB pipeline \citep{Ferrarese12}, and subtracted an elliptical isophotal model of M59. 
Figure~\ref{fig:stream} shows the model-subtracted image. The cross-shaped residual is typical in galaxies that show significant diskiness, such as M59. What we find notable is the broad linear feature to the south of M59, which extends for at least $\sim200\arcsec$, or 14~kpc at the distance of M59. This ``plume'' has $\langle\mu_{g'}\rangle\approx28.0$~\magarcsec, which is 1~mag above the detection threshold for NGVS images \citep[][see Figure~18]{Ferrarese12}. If extended toward the inner regions of M59, the plume roughly lines up with \objname. Although it is possibile the plume is associated with the nearby irregular galaxy IC~3665 (VCC~1890) to the southwest, its existence in the vicinity of M59 and \objname\ suggests a common origin with the larger galaxy.

We mask bright sources and median bin the image by $9\times9$ pixels (following \citealt{Mihos05}), and measure photometry of the ``plume'' along the length of the line shown in Figure~\ref{fig:stream}, avoiding the residuals in the inner region, and using sky regions alongside the feature. The total magnitude is $g'=16.8\pm0.4$~mag, which at M59's distance corresponds to $M_{g'}=-14.2\pm0.4$~mag ($L_{g'}=5.3\times10^7 L_\odot$), and is comparable to \objname\ itself. The quoted uncertainty is for random errors, but we found that selecting different sky regions around the plume, or using different isophotal model subtractions of M59, can change the measured brightness by a few tenths of a magnitude. A similar measurement in the $i'$-band yields $i'=15.7\pm0.2$~mag and $g'-i'=1.1\pm0.4$~mag, which is consistent with the colors of \objname\ and M59, within the uncertainties. 

If associated with \objname, the luminosity of this plume would be a lower-limit on the amount of material stripped from \objname's progenitor. Although significant mass could be hidden at lower surface brightnesses, hiding 1--2 orders of magnitude more mass would require it to be spread over 10--100 times more area. It seems more likely that either \objname\ was a high fraction of its progenitor's total stellar mass, the stripping happened a long time ago, or \objname\ was originally formed compact and dense. Measuring the mass of any central black hole in \objname\ could help differentiate these scenarios.

\section{Conclusions}
\label{sec:conclusions}

We report on the properties of \objname, the most massive UCD in the Virgo cluster and the densest galaxy known, with $M_{g'}=-14.2$~mag and $R_e=25\pm2$~pc. \objname\ has a dynamical mass of $(3.7\pm0.4)\times10^8 M_\odot$, and a mass-to-light ratio of $M_{dyn}/L_{g'}= 7.1\pm0.7$. This $M/L$ is consistent with that expected for an old (14~Gyr), metal-rich ([Fe/H]$=-0.01\pm0.21$, [$\alpha/$Fe]$=+0.21\pm0.07$) stellar population with an IMF intermediate between Kroupa and Salpeter. Given the discovery of a supermassive black hole in M60-UCD1, \objname\ is also likely to harbor one. This object, and nearby M60-UCD1, are likely to be a class of objects that is intermediate between lower mass UCDs and more massive and larger cEs. 

We identify a tidal stream to the south of M59 whose length roughly aligns with the direction of \objname. The measured luminosity of this stream ($M_{g'}=-14.2\pm0.4$~mag), if associated with \objname, provides a lower limit on the luminosity of stars that have been stripped. This relatively small amount of stellar light implies that \objname\ may not be simply a ``naked'' nuclear star cluster, but the remnant of a stripped bulge or pure-\sersic\ galaxy. Future searches for a central black hole and for tidal debris at lower surface brightness levels would help illuminate the nature of \objname\ and the class of objects it represents.

\bigskip
\acknowledgments

The NGVS team owes a debt of gratitude to the director and the staff of the Canada-France-Hawaii Telescope (CFHT), whose dedication, ingenuity, and expertise have helped make the survey a reality. We thank Connie Rockosi for help with the Keck observations.

CL acknowledges the National Key Basic Research Program of China (2015CB857002), NSFC grants 11203017 and 11125313, and from the Office of Science and Technology, Shanghai Municipal Government (11DZ2260700). EWP acknowledges NSFC grants (11173003, 11573002), and the Strategic Priority Research Program of CAS (XDB09000105). NSF support is acknowledged through grants AST-1010039 (PG) and AST-1108964 (JCM).

Based on observations obtained with MegaPrime/MegaCam, a joint project of CFHT and CEA/DAPNIA, at the CFHT which is operated by the National Research Council (NRC) of Canada, the Institut National des Sciences de Univers of the Centre National de la Recherche Scientifique (CNRS) of France, and the University of Hawaii.

This work was supported by the Sino-French LIA-Origins joint exchange program 
and by the Canadian Advanced Network for Astronomical Research (CANFAR) which has been made possible by funding from CANARIE under the Network-Enabled Platforms program. 

{\it Facilities}: CFHT, UH88, Keck


 \newcommand{\noop}[1]{}

\end{document}